\documentclass[english,aps,preprint]{revtex4}
\usepackage[T1]{fontenc}
\usepackage[latin9]{inputenc}
\usepackage{amsmath}
\usepackage{graphicx}
\usepackage{amssymb}

\usepackage{babel}

\begin{document}

\title{Universal protection of unitary evolution from slow noise: dynamical
control pushed to the extreme}

\author{G. Bensky$^{1}$, E. Brion$^{2}$, F. Carlier$^{2}$, V.M. Akulin$^{2}$,
and G. Kurizki$^{1}$}

\affiliation{$^{1}$Weizmann Institute of Science, Department of Chemical Physics,
Rehovot 76100, Israel}

\affiliation{$^{2}$Laboratoire Aimé Cotton, CNRS, Campus d'Orsay, 91405, Orsay,
France}
\begin{abstract}
We propose a technique that allows to simultaneously perform universal
control of the evolution operator and compensate for the first order
contribution of an arbitrary Hermitian constant noise. We show that,
at least, a three-valued Hamiltonian is needed in order to protect
the system against any such noise. This technique is illystrated by
an explicit algorithm for a control sequence that is applied to numerically
design a safe two-qubit gate. 
\end{abstract}
\maketitle

\section{Introduction}

Within the last decades, quantum control has emerged as one the most
fruitful fields in both theoretical and experimental physics \cite{BS90,DA08}.
In particular, it is of crucial importance in quantum computation
\cite{NC00}. To process the information stored in the computer state,
one must indeed be able to generate any prescribed unitary evolution
operator on the computer --- or, at least a universal set of such
operators, unaffected by quantum errors arising from the interaction
with the environment.

Different strategies have been developed to control the evolution
of closed quantum systems, including optimal control approaches \cite{PDR88,PK02,SGL99,OKF98,SR90}
and algebraic methods \cite{RFRO00,RSDRP95,SGRR02,ZR99}. To deal
with open quantum systems, schemes have been designed to correct/avoid
the undesired effects due to the environment. Quantum error-correcting
codes (QEC) \cite{S95,NC00,KL97,G96,CS96} and approaches based on
the quantum Zeno effect \cite{BAC05,BHK04} use redundancy of encoding
as a way to recover information after the errors occur. Topological
protection \cite{IFIITB02} takes advantage of the symmetries of the
system to safely store information in so-called Decoherence Free Subspaces
\cite{LBKW01,LBKW01b,KBLW01}. An alternative to QEC that is substantially
less resource-intensive is dynamical decoupling (DD) \cite{VZKTDD,KL05,Uhrig07}.
In DD one applies a succession of short and strong pulses to the system,
designed to stroboscopically decouple it from the environment. Similar
in spirit to DD, but more general, is the method we term here {}``dynamical
control by modulation'' (DCM), wherein one may apply to the system
a sequence of arbitrarily-shaped pulses whose duration may vary anywhere
from the stroboscopic limit to that of continuous dynamical modulation
\cite{BAC05,KK05,Agarwal99,Alicki04,Facchi05,GKK06}. In the DCM approach,
the decoherence rate is governed by a universal expression, in the
form of an overlap between the bath-response and modulation spectra.
In such methods, it is however not clear a priori whether one can,
at the same time, protect the system from noise and perform \emph{universal}
control of its evolution operator.

In this paper we investigate this question by asking whether it is
possible to perform any arbitrarily chosen evolution of a quantum
system while compensating for all Hermitian static noises. To be more
specific, our goal here is to show how to design a time-dependent
control Hamiltonian which can, at the same time, impose a chosen evolution
to the system and eliminate the first order action of any Hermitian
static noise. We show that, contrary to strict evolution control problems,
this objective cannot be achieved with only two-valued Hamiltonians
but requires the use of at least three control operators. We go on
to show that even a {\em null} third operator, causing the system
to evolve under the action of noise only, is enough. Inspired by a
previous result \cite{HA99}, we propose a new algorithm able to compute
the appropriate control sequence for any given desired evolution.
As an application, this algorithm was run to design a safe $CNOT$
gate.

The paper is structured as follows. We first set the problem to solve
and give the explicit conditions the evolution matrix must fulfill.
Then we show that these conditions are impossible to meet by a two-valued
control Hamiltonian. Slightly modifying the two-stage procedure by
allowing for extra steps during which the control Hamiltonian is set
to zero, we describe an algorithm able to compute an appropriate protected
control sequence. An application to a two-qubit gate is finally proposed.

\section{Conditions for first-order noise elimination}

Let us consider an $N$-level quantum system whose Hamiltonian consists
of a \emph{controllable} part, denoted by $H_{c}\left(t\right)$,
and an \emph{unknown} but \emph{static} noise contribution, which
can be written as a linear combination $\mathcal{N}=\sum_{i}\hbar\varepsilon_{i}G_{i}$
of the Hermitian traceless generators $\left\{ G_{i},i=1,\ldots,N^{2}-1\right\} $
of $su(N)$. The evolution matrix satisfies the dynamical equation
\begin{eqnarray}
\imath\hbar\frac{\partial U\left(t\right)}{\partial t} & = & \left(H_{c}\left(t\right)+\mathcal{N}\right)U\left(t\right)\\
U\left(0\right) & = & I.\end{eqnarray}
 Upon transforming to the interaction picture relative to $H_{c}$,
one isolates the evolution $\widetilde{U}\left(t\right)$ due to the
noise only: defining \begin{equation}
U_{c}\left(t\right)\equiv T\exp\left\{ \frac{1}{\imath\hbar}\int_{0}^{t}H_{c}\left(s\right)ds\right\} \end{equation}
 the evolution induced by the control Hamiltonian alone, where $T$
denotes the chronological product, one sets $\widetilde{U}\left(t\right)\equiv U_{c}^{\dagger}\left(t\right)U\left(t\right)$,
which satisfies \begin{eqnarray}
\imath\hbar\frac{\partial\widetilde{U}}{\partial t} & = & \left[U_{c}^{\dagger}\left(t\right)\mathcal{N}U_{c}\left(t\right)\right]\widetilde{U}\left(t\right)\\
 & = & \left[\sum_{i}\hbar\varepsilon_{i}U_{c}^{\dagger}\left(t\right)G_{i}U_{c}\left(t\right)\right]\widetilde{U}\left(t\right)\\
\widetilde{U}\left(0\right) & = & I.\end{eqnarray}
 The first order contribution of the noise to the evolution is thus
given by the second term in the Dyson expansion of $\widetilde{U}\left(t\right)$
for the accumulated action, that is \begin{equation}
\widetilde{U}^{\left(1\right)}\left(t\right)=\sum_{i}\varepsilon_{i}\int_{0}^{t}U_{c}^{\dagger}\left(s\right)G_{i}U_{c}\left(s\right)ds.\end{equation}

Our goal is to design a control Hamiltonian $H_{c}\left(t\right)$,
such that, at the end of the control sequence, say at time $T_{c}$,
the evolution operator takes an arbitrarily prescribed value $U_{d}\in SU\left(N\right)$
while the first order contribution of any constant noise vanishes.
We thus require \begin{eqnarray}
U_{c}\left(T_{c}\right) & = & U_{d}\label{Condition1}\end{eqnarray}
 and, for any set of constants $\left\{ \varepsilon_{i}\right\} $,
\begin{equation}
\widetilde{U}^{\left(1\right)}\left(T_{c}\right)=\sum_{i}\varepsilon_{i}\int_{0}^{T_{c}}U_{c}^{\dagger}\left(s\right)G_{i}U_{c}\left(s\right)ds=0,\end{equation}
 that is \begin{equation}
\forall i,\quad\int_{0}^{T_{c}}U_{c}^{\dagger}(s)G_{i}U_{c}(s)ds=0.\label{Condition2}\end{equation}

\section{Alternating two-valued operator sequence}

Let us now focus on the two-valued alternating perturbation approach.
Namely, the control Hamiltonian $H_{c}\left(t\right)$ alternates
between two values $A$ and $B$ for adjustable timings which play
the role of control parameters. Formally, $H_{c}\left(t\right)$ then
assumes the bilinear form $H_{c}\left(t\right)=\hbar\alpha\left(t\right)A+\hbar\beta\left(t\right)B$,
where $\alpha\left(t\right)$ and $\beta\left(t\right)$ are two piecewise
constant functions taking the values $0,1$ and adding up to $1$.
In the sudden approximation, the overall evolution operator induced
by such a $K$-step control sequence has the pulsed form \begin{equation}
U_{c}\left(T_{c}\right)=e^{-\imath T_{K}H_{K}}\times\ldots\times e^{-\imath T_{2}H_{2}}\times e^{-\imath T_{1}H_{1}},\end{equation}
 where $H_{k}\equiv A$ when $k$ is even, $B$ when $k$ is odd,
and $\sum_{k=1}^{K}T_{k}=T_{c}$.

Provided that $A$, $B$ together with their all-order commutators
span $su(N)$ (the bracket generation condition), it can be shown
\cite{HA99} that it is possible to design $K\propto N^{2}$ such
control timings $\left\{ T_{1},\ldots,T_{K}\right\} $ for which eq.(\ref{Condition1})
is met, \emph{i.e.} such that \begin{equation}
e^{-\imath T_{K}H_{K}}\times\ldots\times e^{-\imath T_{2}A}\times e^{-\imath T_{1}B}=U_{d}.\end{equation}

It turns out, however, that, except for the particular case $U_{d}=I$,
eq.(\ref{Condition2}) cannot be satisfied for all noises. As we shall
now show, there indeed always exists a $D$-dimensional uncorrectable
subspace $\left(1\leq D\leq N-1\right)$ spanned by the noise operators
$C_{n\in N}\equiv\frac{1}{2}\left[A+B,\left(B-A\right)^{n}\right]$.
To prove this, let us first introduce the operators \begin{equation}
U_{0}\equiv I,\quad U_{K\geq i\geq1}\equiv e^{-\imath T_{i}H_{i}}\times\ldots\times e^{-\imath T_{2}A}\times e^{-\imath T_{1}B}\label{Udefinition}\end{equation}
 which allow us to write \begin{eqnarray}
U_{c}\left(t\right) & = & e^{-\imath\left(t-\sum_{k=1}^{i}T_{k}\right)H_{i+1}}U_{i}\\
 &  & (\text{for }{\textstyle \sum_{1}^{i}T_{k}\leq t<\sum_{1}^{i+1}T_{k}}).\label{UCdefinition}\end{eqnarray}
 Let us further define the superoperators $\hat{\mathcal{H}}_{i}$,
whose action is given, on any operator $X$, by $\hat{\mathcal{H}}_{i}X\equiv\left[H_{i},X\right]$.
We can thus write $e^{+\varepsilon H_{i}}Xe^{-\varepsilon H_{i}}=\sum_{k=0}^{\infty}\frac{\varepsilon^{k}}{k!}\hat{\mathcal{H}}_{i}^{k}X$.
Since $\left[A,(B-A)^{n}\right]=\left[B,(B-A)^{n}\right]=\left[H_{i},(B-A)^{n}\right]$
for every $i$, we can set $C_{n}=\hat{\mathcal{H}}_{i}(B-A)^{n}$.

Let us now evaluate the first order effect of $C_{n}$ according to
eq.(\ref{Condition2}): \begin{eqnarray}
\int_{0}^{T_{c}} & U_{c}^{\dagger}(s)C_{n}U_{c}(s)ds\\
 & = & \sum_{i}\int_{0}^{T_{i}}dsU_{i}^{\dagger}e^{\imath sH_{i}}\hat{\mathcal{H}}_{i}(B-A)^{n}e^{-\imath sH_{i}}U_{i}\\
 & = & \sum_{i}U_{i}^{\dagger}\int_{0}^{T_{i}}ds\sum_{k=0}^{\infty}\frac{(-\imath s)^{k}}{k!}\hat{\mathcal{H}}_{i}^{k+1}(B-A)^{n}U_{i}\\
 & = & \imath\sum_{i}U_{i}^{\dagger}\left[e^{\imath T_{i}H_{i}}(B-A)^{n}e^{-\imath T_{i}H_{i}}-(B-A)^{n}\right]U_{i}\\
 & = & \imath\sum_{i}U_{i+1}^{\dagger}(B-A)^{n}U_{i+1}-\imath\sum_{i}U_{i}^{\dagger}(B-A)^{n}U_{i}\\
 & = & \imath U^{\dagger}(T_{c})(B-A)^{n}U(T_{c})-\imath(B-A)^{n}.\end{eqnarray}
 The first order contribution we have just obtained does not depend
on the specific timing parameters $T_{i}$'s but only on the final
operation $U(T_{c})$; it is moreover nonzero for any $U_{d}\neq I$.
Finally, according to the Cayley-Hamilton theorem, the matrix $(B-A)$
cancels its characteristic polynomial, which implies that $1\leq\dim\left[\mathrm{Span}\left(\left\{ (B-A)^{n},n\in N\right\} \right)\right]\leq N-1$.
As a consequence the dimension $D\equiv\dim\left[\mathrm{Span}\left(\left\{ C_{n},n\in N\right\} \right)\right]$
of the uncorrectable subspace satisfies $1\leq D\leq N-1$.

Let us now examine how the two-operator Hamiltonian scheme can be
modified so that {\em both} conditions eq.(\ref{Condition1},\ref{Condition2})
are met. Suppose we have a sequence of timings $\{t_{1},\dots,t_{K}\}$.
Now $U_{0}$ and $U_{n\geq1}$ are the evolution operator at the end
of the $n^{th}$ step of the control sequence as defined in eq.(\ref{Udefinition}),
while $U_{c}(t)$ is the evolution operator at time $t$ as defined
in eq.(\ref{UCdefinition}). Following the method described in \cite{HA99},
the last $N^{2}-1$ timings $\{t_{K-N^{2}+2},\dots,T_{K}\}$ can be
chosen such that they achieve the evolution $U_{d}\cdot U_{K-N^{2}+1}^{\dagger}$,
thus satisfying the condition of eq.(\ref{Condition1}). The first
order contribution of any noise $G_{i}$ can be explicitly expressed
by the accumulated action \begin{equation}
\mathcal{G}_{i}\equiv\sum_{n=1}^{K}U_{n}^{\dagger}g_{i,n}U_{n},\end{equation}
 where $g_{i,n}\equiv\int_{0}^{T_{n}}e^{-\imath sH_{n}}G_{i}e^{+\imath sH_{n}}ds$.

Let us, at each commutation between A and B, allow for the waiting
time $\tau_{l=1,\dots,K}$ during which no perturbation is applied.
This amounts to adding a \emph{third} value $C=0$ to the control
Hamiltonian $H_{c}(t)$. We see that the overall evolution operator
remains unchanged, while the first order contribution of any noise
$G_{i}$ is added the term $\sum_{n=1}^{K}\tau_{n}U_{n}^{\dagger}G_{i}U_{n}$,
which is a linear function of the waiting times.

Let us choose the timings $\{t_{1},\dots,t_{K-N^{2}+1}\}$ such that
the coefficients of the waiting times span the entire $N^{2}\cdot(N^{2}-1)$
space. Generally this can be achieved by randomly choosing the timings.
Thus, by solving a simple set of $N^{2}\cdot(N^{2}-1)$ linear equations
of the form $\mathcal{G}_{i}+\sum_{n=1}^{K}\tau_{n}\mathcal{F}_{i,n}=0$
($\mathcal{F}_{i,n}\equiv U_{n}^{\dagger}G_{i}U_{n}$) to find the
waiting times $\tau_{1},\dots,\tau_{K}$, one can eliminate the first
order contribution of all the noises $G_{i}$ added during the $A$,
$B$ control sequences. In practice, since the waiting times $\tau_{i}$
must be positive, this requires a slightly larger $K$ and the use
of linear programming methods or other minimization techniques.

\section{Algorithm for a four-state system}

\begin{figure}
\begin{centering}
\includegraphics[width=0.9\linewidth]{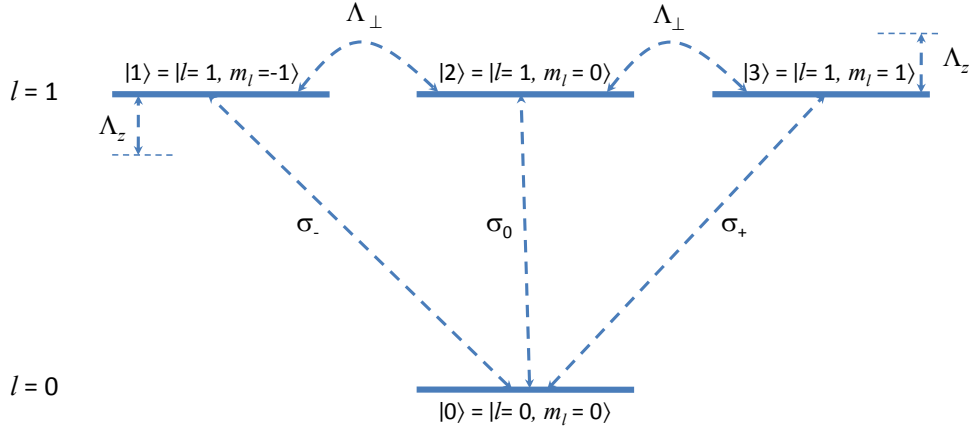} 
\par\end{centering}

\caption{ The two-level atomic system ($l=0,1$). The arrows show the different
couplings due to the resonant electric field (operators $\sigma_{-,0,+}$)
and the static magnetic field (operators $\Lambda_{\perp,z}$).}

\label{Fig} 
\end{figure}

This algorithm was applied to a model four-state system, represented
in Fig. \ref{Fig}, which can be used to store and process two qubits
of information. The four states correspond to two different angular
momenta $l=0,1$: $\left|0\right\rangle \equiv\left|l=0,m_{l}=0\right\rangle $,
$\left|1\right\rangle \equiv\left|l=1,m_{l}=-1\right\rangle $, $\left|2\right\rangle \equiv\left|l=1,m_{l}=0\right\rangle $
and $\left|3\right\rangle \equiv\left|l=1,m_{l}=1\right\rangle $.

This system is subject to a resonant electric and a static magnetic
fields: the $\pi$-component of the electric field couples $\left|0\right\rangle $
to $\left|2\right\rangle $ while the $\sigma_{+,-}$-components couple
$\left|0\right\rangle $ to $\left|1\right\rangle $, and $\left|0\right\rangle $
to $\left|3\right\rangle $, respectively; the $x,y$-component of
the magnetic field couples $\left|1\right\rangle $ to $\left|2\right\rangle $
and $\left|2\right\rangle $ to $\left|3\right\rangle $ while its
$z$-component shifts $\left|1\right\rangle $ and $\left|3\right\rangle $
out of resonance. Finally, in the rotating wave approximation (RWA),
the total Hamiltonian of the system assumes the form \begin{eqnarray}
H_{c}= & e_{-}\sigma_{-}+e_{0}\sigma_{0}+e_{+}\sigma_{+}+b_{\perp}\Lambda_{\perp}+b_{z}\Lambda_{z}\\
\sigma_{-}\equiv & \left|1\right\rangle \left\langle 0\right|+\left|0\right\rangle \left\langle 1\right|\\
\sigma_{0}\equiv & \left|2\right\rangle \left\langle 0\right|+\left|0\right\rangle \left\langle 2\right|\\
\sigma_{+}\equiv & \left|3\right\rangle \left\langle 0\right|+\left|0\right\rangle \left\langle 3\right|\\
\Lambda_{\perp}\equiv & \left|1\right\rangle \left\langle 2\right|+\left|2\right\rangle \left\langle 3\right|+\left|2\right\rangle \left\langle 1\right|+\left|3\right\rangle \left\langle 2\right|\\
\Lambda_{z}\equiv & \left|3\right\rangle \left\langle 3\right|-\left|1\right\rangle \left\langle 1\right|\end{eqnarray}
 where $e_{-,0,+}$ and $b_{\perp,z}$ are five independent parameters,
proportional to the electric and magnetic field amplitudes, respectively.
By choosing two different sets $\left\{ e_{-,0,+}^{A},b_{\perp,z}^{A}\right\} $
and $\left\{ e_{-,0,+}^{B},b_{\perp,z}^{B}\right\} $ of such parameters,
we can define two values for the Hamiltonian $H_{c}=A,B$ which may
be used as our alternating perturbations.

After verifying that $A$ and $B$ satisfy the bracket generation
condition, we first calculate the $K=N^{2}=16$ timings $\left\{ t_{k=1,\ldots,K}^{\left(0\right)}\right\} $
which realize the identity matrix, by the method described in \cite{HA99},
based on statistical properties of the roots of the identity. By the
Newton iterative method, we then compute the $K=16$ timings $\left\{ t_{k=1,\ldots,K}\right\} $
which perform the transformation $CNOT^{\frac{1}{N^{2}-1}}=CNOT^{\frac{1}{15}}$,
where $CNOT\equiv\left(\begin{array}{cccc}
1 & 0 & 0 & 0\\
0 & 1 & 0 & 0\\
0 & 0 & 0 & 1\\
0 & 0 & 1 & 0\end{array}\right)$.

Finally, we look for the $K\left(N^{2}-1\right)=240$ waiting times
$\tau_{l=1,\ldots,K\left(N^{2}-1\right)}$ which minimize all $15$
possible noise operators to zero. By applying the complete $480$-step
control sequence, consisting of the $\left(N^{2}-1\right)=15$ repetitions
of the sequence $\left\{ t_{k=1,\ldots,K=16}\right\} $ plus the $K\left(N^{2}-1\right)=240$
waiting periods $\left\{ \tau_{l}\right\} $ with zero control after
each step, we can thus impose a safe $CNOT$ gate on the system in
the presence of \emph{any} noise that varies slower than the control
sequence.

\section{Discussion}

It is plausible that the conditions eq.(\ref{Condition1},\ref{Condition2})
can be satisfied by an arbitrary three-valued Hamiltonian, with $C\neq0$.
We have looked for an appropriate sequence of timings with such a
Hamiltonian by direct optimization of a cycle of $KN^{2}$ operations
and found satisfactory numerical solutions up to $N=16$. This approach
is however slower than the scheme described above.

There are several open issues regarding the universal control method
proposed here. First, it is important to treat higher orders in the
perturbation expansion of the noise and, in particular, explore the
feasibility of a control Hamiltonian canceling the second order contribution
of the noise, $i.e.$ the integrals \begin{equation}
\int_{0}^{T_{c}}dt\int_{0}^{t}dsU_{c}^{\dagger}(t)G_{m}U_{c}(t)U_{c}^{\dagger}(s)G_{n}U_{c}(s)\end{equation}
 for any $\left(m,n\right)$. Second, the issue of time-dependent
noises is important. The present method holds for noise that slowly
varies with time, compared to the control sequence. The approach must
change altogether if fast noises affect the system. Finally, it is
imperative to establish whether redundancy can be combined with dynamic
control techniques to safely process the information, as, for example,
in an ensemble of identical systems.

To conclude, we have raised the question whether universal control
of evolution may be performed while compensating for the first order
contribution of arbitrary constant Hermitian noise, by means of an
alternating perturbation procedure. We have demonstrated that this
can not be achieved by a two-valued control Hamiltonian: in that case
there always exists a subspace of uncorrectable noises. If, however,
we allow for waiting times, during which the system is only subject
to noise, our objective becomes feasible. This has been demonstrated
by an explicit algorithm that yields the appropriate control sequence,
as tested on the case $N=4$. The total waiting time is of the order
of half the control period. More generally, we numerically checked
that a three-valued Hamiltonian can control the evolution operator
and protect it against any static Hermitian noise.
\begin{acknowledgments}

\end{acknowledgments}
G.K. acknowledges the support of the EC (SCALA and MIDAS), GIF and
ISF.

\end{document}